\documentclass[cameraready]{Interspeech}

\title{PolyBench: A Benchmark for Compositional Reasoning in Polyphonic Audio}
\author[affiliation={1},equalcontribution]{Yuanjian}{Chen}
\author[affiliation={2},equalcontribution]{Yang}{Xiao}
\author[affiliation={3},equalcontribution]{Han}{Yin}
\author[affiliation={4}]{Xubo}{Liu}
\author[affiliation={1}]{Jinjie}{Huang}
\author[affiliation={2}]{Ting}{Dang}

\address{
    $^1$ Harbin University of Science and Technology, China \\
    $^2$ The University of Melbourne, Australia \\
    $^3$ KAIST, South Korea \\
    $^4$ University of Surrey, UK
}

\email{2010400002@stu.hrbust.edu.cn, yxiao9550@student.unimelb.edu.au, hanyin@kaist.ac.kr}

\usepackage{soul, xcolor}

\definecolor{myblue}{RGB}{200,220,235}

\usepackage{comment}
\usepackage{booktabs} 
\usepackage{amssymb}  

\begin{document}

\maketitle

\begin{abstract}
Large Audio Language Models (LALMs) are increasingly capable of reasoning over audio. However, existing benchmarks provide limited coverage of reasoning in polyphonic audio, where multiple sound events co-occur and induce compositional structure. In this work, we introduce PolyBench\footnote{https://huggingface.co/datasets/PolyBench/PolyBench}, a benchmark designed to evaluate compositional reasoning in polyphonic audio. PolyBench comprises five evaluation subsets covering counting, classification, detection, concurrency, and duration estimation, requiring reasoning over multiple concurrent events and their relations. Evaluation of state-of-the-art LALMs reveals consistent performance degradation in polyphonic audio, indicating a fundamental bottleneck in current LALMs. 

\end{abstract}

\section{Introduction}
Large Language Models are reshaping the paradigm of audio perception, giving rise to Large Audio Language Models (LALMs)~\cite{chu2023qwen,su2025audio,xie2025audio,luo2026survey} with strong understanding and reasoning capabilities. Earlier generations of audio models were primarily designed for closed-set recognition tasks, such as sound classification or event detection under relatively clean conditions~\cite{kong2020panns,yin2025exploring,chen2025noise,yin2026focus}. In contrast, modern AI systems are expected to operate in open and dynamic acoustic environments, where multiple sound sources interact, overlap, and evolve over time. This shift introduces fundamentally new challenges: beyond identifying individual events, models must reason about temporal structure, compositional relationships, and cross-event dependencies. Such capabilities are critical for emerging applications, including audio question answering~\cite{lipping2022clotho} and multimodal embodied agents~\cite{szot2025multimodal}.

Recent large-scale efforts have made significant progress in developing reasoning benchmarks for Large Audio Language Models (LALMs)~\cite{sakshimmau,mammar,yuan2024t,yin2025envsdd}. For example, AIR-Bench~\cite{yang2024air} evaluates how well LALMs follow instructions across diverse audio signals. Meanwhile, MMAU-Pro~\cite{kumar2025mmaup} provides a comprehensive evaluation of 49 auditory skills, with a particular emphasis on long-form audio understanding. Despite these advances, evaluations on MMAU-Pro and related benchmarks reveal that state-of-the-art models still struggle with temporal event reasoning. To address this challenge, benchmarks such as TREA~\cite{tera} introduce datasets specifically designed for fine-grained temporal reasoning.

Despite the progress made by existing benchmarks, systematic evaluation of polyphony remains limited. Most benchmarks primarily assess the chronological ordering or duration of isolated, sequential events, and rarely consider scenarios where multiple sound sources overlap simultaneously~\cite{cakir2017polyphonic,xiao2024wilddesed,xiao2025dg}. Polyphony introduces unique challenges beyond standard temporal reasoning: overlapping sources make it difficult to distinguish individual sound events, while simultaneously inducing compositional structures that require reasoning about relationships among concurrent events. Although some recent benchmarks incorporate audio mixtures~\cite{mammar}, they do not explicitly evaluate compositional reasoning in polyphonic audio scenes.

In this work, we introduce PolyBench, the first benchmark designed to evaluate compositional reasoning in polyphonic audio. To match realistic acoustic scenarios, we construct PolyBench from polyphonic audio clips sampled from real-world recordings, including three sources: DataSED \cite{DataSED}, DESED \cite{DESED}, and MAESTRO-Real \cite{maestro}. PolyBench comprises five evaluation subsets covering counting, classification, detection, concurrency, and duration estimation, each requiring reasoning over multiple concurrent audio events and their relationships. Our evaluation shows that state-of-the-art LALMs consistently struggle in polyphonic settings, exposing a key bottleneck in compositional reasoning over complex acoustic scenes.

\begin{figure*}[t]
\centering
\includegraphics[width=1\textwidth]{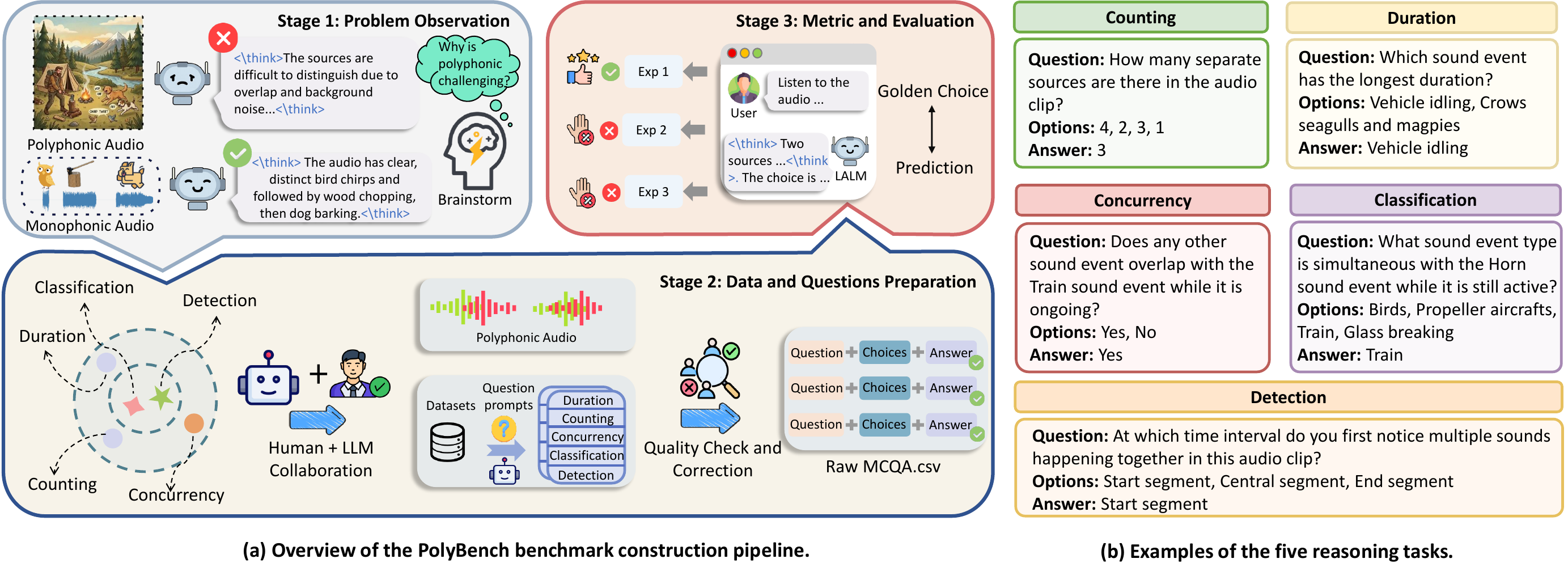}
\vspace{-16pt}
\caption{PolyBench pipeline for benchmark construction and evaluation. The process includes (1) observing polyphony-induced failure patterns by contrasting monophonic and polyphonic audio; (2) curating real-world polyphonic clips and generating MCQAs for five task types via human–LLM collaboration with iterative quality control. The circles denote the five tasks, which are from basic perception to complex reasoning; and (3) evaluating LALMs by scoring their MCQA predictions against golden answers. The right panel shows representative examples of the five tasks (Counting, Duration, Concurrency, Classification, and Detection).
}
\vspace{-12pt}
\label{fig:fig1}
\end{figure*}

\section{ Benchmark}


As illustrated in Figure~\ref{fig:fig1}, the construction and evaluation pipeline of our benchmark consists of three primary stages: In stage 1, we focus on problem observation. In stage 2, we involve data curation and question generation, and stage 3 encompasses quantitative measurement and systematic evaluation.

\subsection{Stage 1: Problem Observation}


During our preliminary investigation of audio captioning tasks, we found that LALMs can accurately identify distinct sound event categories and infer their temporal relationships when processing non-overlapping monophonic audio. As illustrated in Stage 1 of Figure~\ref{fig:fig1}, existing LALMs exhibit robust recognition and reasoning capabilities for monophonic audio understanding \cite{wang2025timeaudio,kumar2026tac}. However, when multiple sound events occur simultaneously, resulting in polyphonic audio, these models often suffer from substantial confusion and hallucination, leading to ineffective identification of concurrent sound sources. Motivated by this limitation, we define the primary objective of this work as follows: \textbf{to develop a comprehensive audio understanding benchmark specifically designed for overlapping sound events}.

\subsection{Stage 2: Data and Questions Preparation}
Multiple-Choice Question Answering (MCQA) has been widely adopted as the standard evaluation paradigm for assessing the capabilities of LALMs. 
To evaluate the performance of LALMs in polyphonic audio understanding, PolyBench introduces five categories of MCQA tasks that are systematically organized according to a concentric hierarchical structure of polyphonic reasoning dependencies. These tasks are designed to provide a comprehensive and quantitative assessment of temporal understanding and reasoning capabilities with respect to overlapping sound events.

\begin{figure}[!t]
  \centering
  \includegraphics[width=\linewidth]{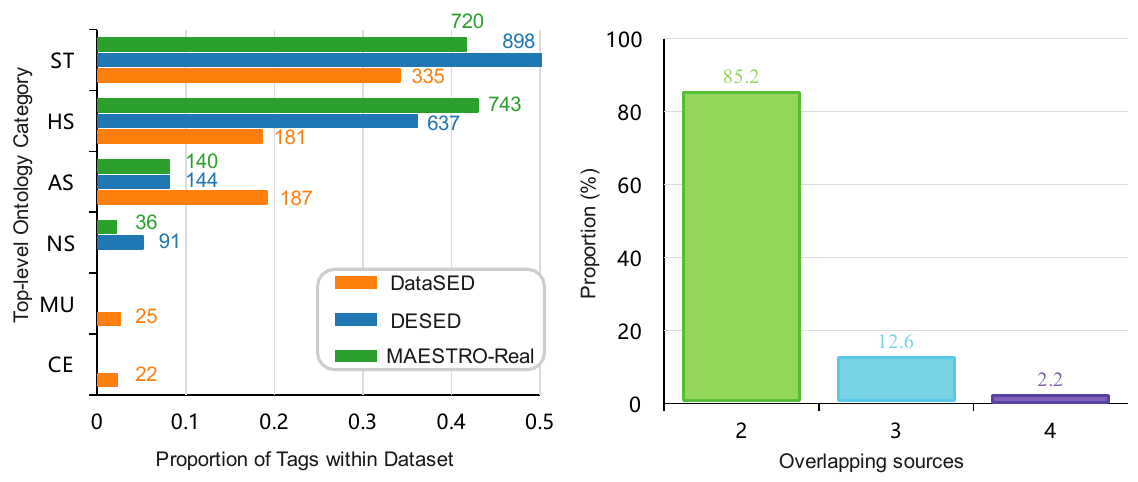}
  \vspace{-16pt}
  \caption{Visual analysis of PolyBench dataset characteristics. Left: distribution of PolyBench event tags across top-level AudioSet ontology categories—CE (Channel, environment and background), MU (Music), NS (Natural sounds), AS (Animal sounds), HS (Human sounds), and ST (Sounds of things). Numbers at the end of each bar indicate the event occurrence counts in the dataset, while the x-axis shows the proportion of tags within each source dataset. Right: proportion of audio clips by the number of overlapping sound sources (e.g., 2, 3, and 4 concurrent events).}
  \label{fig:fig2}
  \vspace{-16pt}
\end{figure}

\subsubsection{MCQA Tasks in PolyBench}
We specifically establish a fundamental question as the primary prompt for each task category. Subsequently, we employ Qwen3-Max to produce 20 semantically equivalent variants, which are subsequently allocated to various audio samples at random. The precise definitions and designs of these five task categories are as follows:

\begin{itemize}
    \item \textbf{Counting:} Aim to identify the number of distinct sound event categories in the input audio. Example: \textit{``How many distinct sources does the audio file contain?"}
    \item \textbf{Duration:} Aim to identify the sound event category with the longest duration within the audio. Example: \textit{``Which sound event has the longest duration?"}
    \item \textbf{Concurrency:} Designed to determine whether polyphonic phenomena occur within the audio clip. Example: \textit{``While the frying sound event has not ceased, is there a simultaneous presence of another sound event?"} We ensure that the anchor event mentioned in the question genuinely overlaps temporally with other event categories in the audio.
    \item \textbf{Classification:} Require LALMs to identify the category of another sound event that overlaps with a specified event. Example: \textit{``What type of sound event overlaps temporally with the dog sound event before it ends?"} When designing incompatible options, to prevent semantic relevance from interfering with the evaluation of the model's polyphonic parsing ability, we strictly select event categories from the polyphonic subset that are semantically unrelated and do not occur in the current audio clip as perturbation.
    \item \textbf{Detection:} Aim to comprehensively measure the understanding of LALMs of polyphonic phenomena and their temporal reasoning capabilities. Example: \textit{``At what moment in this audio clip do multiple sounds first coincide?"} Unlike the sound event detection (SED) task, this task does not require LALMs to output precise timestamps. Instead, it focuses on perceiving the temporal intervals where overlapping events occur and implicitly reasoning about the total number of overlap occurrences.
\end{itemize}

After generating the five specified question types, we perform thorough manual quality check and corrections to ensure logical coherence among the ``questions-choices-answer" triplets. The completed triplets are stored as a CSV file that functions as the input prompts for LALMs in the next stage.

\subsubsection{Audio Sources and Statistics}

To match realistic acoustic scenarios, we sample polyphonic audio clips from real-world recordings, including three sources: DataSED \cite{DataSED}, DESED \cite{DESED} and MAESTRO-Real \cite{maestro}.

Specifically, \noindent \textbf{DataSED} dataset contains 4,292 real-world audio samples covering 22 sound event classes. We extract audio segments containing overlapping sound events. Finally, 169 clips are selected for inclusion in PolyBench, with sample durations ranging from 3 seconds to 89 seconds.
\noindent \textbf{DESED} is a widely used dataset for SED in home scenario. We specifically sample audio clips exhibiting polyphonic phenomena from its official evaluation set, yielding a total of 259 samples.
\noindent \textbf{MAESTRO-Real} is an SED dataset in various indoor and outdoor environments. We extract audio clips containing overlapping events from it. To balance the overall duration distribution of the samples within PolyBench, we uniformly clip long audio samples in MAESTRO-Real to 25 seconds audio clips and finally get 300 samples. For all samples obtained from above sources, we have exact ground truth of sound event classes and corresponding timestamps.
Figure \ref{fig:fig2} presents the statistical distribution of sound events in PolyBench.

It should be noted that for the \textbf{Concurrency} task, if only polyphonic audio clips were included, the correct answer would always be ``\textit{Yes}'', since overlapping events are inherently present, leading to a trivial bias and weakening the evaluation validity. To avoid this issue, we additionally sample monophonic audio clips without any temporal overlap from AudioTime~\cite{AudioTime}, ensuring a balanced distribution of positive and negative samples and enabling a more reliable assessment of models’ ability to perceive concurrent sound events.


\subsection{Stage 3: Metric and Evaluation}

In this stage, we establish a standardized evaluation paradigm, employing an MCQA format to evaluate the LALMs. During the evaluation process, the LALM receives an audio input alongside a textual instruction generated in the Stage 2. For reasoning models, to activate the models' deep reasoning capabilities for the polyphonic condition, we prompt them to generate their internal reasoning processes (i.e., Chain-of-Thought, CoT) \cite{kumar2025mmaup, tera, chainofthought} prior to outputting their final choice. The final quantitative metrics are calculated strictly based on the exact match between the model's final selected option and the answer within the triplets.
Following MMAU-Pro, we use the average accuracy (\textbf{ACC}) and \textbf{F1} score as the objective metrics.


Furthermore, to better characterize the behavior of reasoning-based LALMs, we conduct a fine-grained analysis based on TP, FP, TN, and FN cases. Specifically, each prediction is compared with the ground-truth answer and categorized as TP, FP, TN, or FN. This protocol complements aggregate metrics by revealing whether performance limitations arise from false alarms, missed detections, or incorrect temporal reasoning over overlapping sound events.

\section{Experimental Setups}

In this benchmark, we select a diverse array of highly influential, fully open-source, and well-documented LALMs released in the past year as our baselines. These specifically encompass the Audio Flamingo 3 (Large Audio Language Model, LALM) \cite{flamingo3}, R1-AQA (Large Audio Reasoning Model, LARM) \cite{R1AQA}, and Qwen3-Omni-30B-A3B (Omni Model) \cite{qwen3Omni}. We also evaluate a cascaded system consisting of TimeAudio \cite{wang2025timeaudio} and Qwen3-8B \cite{qwen3}, this system designates TimeAudio to temporally localize sound events and produce corresponding captions, which are subsequently analyzed using the sophisticated text reasoning capabilities of Qwen3-8B. Furthermore, we integrate AUDSEMTHINKER-QA GRPO \cite{audsemthinker}, a model explicitly developed for semantic reasoning in audio question answering.

To maintain consistency in model evaluation, we utilize explicit prompts to direct the models to produce standard selections from the given choice list. 
In accordance with the experimental setting MMAU-Pro, we employ NV-Embed-v2 to assess the semantic similarity between the model's generated response and the answer. This method effectively reduces the assessment inaccuracies commonly associated with conventional exact string matching or regular expression extraction.

\section{Result and Analysis}

\begin{table*}[t]
\centering
\caption{Performance (\%) of various LALMs on PolyBench compositional reasoning tasks under polyphonic audio (MCQA setting). Best results are in \textbf{bold} and second-best results are \underline{underlined} for each metric column.}
\vspace{-12pt}
\label{tab:polybench_results}
\begin{tabular}{lcccccccccc}
\toprule
\textbf{Models} &
\multicolumn{2}{c}{\textbf{Counting}} &
\multicolumn{2}{c}{\textbf{Duration}} &
\multicolumn{2}{c}{\textbf{Concurrency}} &
\multicolumn{2}{c}{\textbf{Classification}} &
\multicolumn{2}{c}{\textbf{Detection}} \\
\cmidrule(lr){2-3}\cmidrule(lr){4-5}\cmidrule(lr){6-7}\cmidrule(lr){8-9}\cmidrule(lr){10-11}
& \textbf{F1} & \textbf{ACC} & \textbf{F1} & \textbf{ACC} & \textbf{F1} & \textbf{ACC} & \textbf{F1} & \textbf{ACC} & \textbf{F1} & \textbf{ACC} \\
\midrule
Qwen3-Omni-30B-A3B \cite{qwen3Omni}  & 55.9 & \textbf{57.5} & \textbf{70.2} & \textbf{68.5} & 85.2 & 76.0 & \textbf{77.2} & \textbf{77.9} & \textbf{57.3} & \textbf{63.4} \\
R1AQA \cite{R1AQA}             & 34.5 & 30.1          & 42.1          & 43.2          & \textbf{93.7}          & \textbf{90.4}          & 29.5          & 33.1          & 22.4          & 26.2          \\
Audio Flamingo 3 \cite{flamingo3}   & \underline{56.3} & \underline{53.4} & \underline{67.4} & \underline{65.8} & 53.1 & 37.7 & 47.8 & 50.3 & 36.2 & 37.2 \\
TimeAudio \cite{wang2025timeaudio} +Qwen3-8B \cite{qwen3} & 47.7 & 43.8          & 63.7          & 63.7          & \underline{86.9}          & \underline{78.1}          & 53.8          & 50.3          & \underline{53.5} & \underline{51.7} \\
AUDSEMTHINKER-QA GRPO \cite{audsemthinker}     & \textbf{57.3} & 51.4  & 67.1 & 64.4 & 59.1 & 43.2 & \underline{70.6} & \underline{70.3} & 39.6 & 37.2 \\
\bottomrule
\end{tabular}
\vspace{-12pt}
\end{table*}

\begin{table}[t]
\centering
\caption{Performance (\%) of various LALMs on the PolyBench \textbf{Concurrency} task. Best results are in \textbf{bold} and second-best results are \underline{underlined} for each metric column.}
\vspace{-12pt}
\label{tab:concurrency_mixed_mono_confusion}
\setlength{\tabcolsep}{4pt} 
\renewcommand{\arraystretch}{1.05}
\resizebox{\columnwidth}{!}{%
\begin{tabular}{lcccc}
\toprule
\textbf{Model} & \textbf{TP} & \textbf{FP} & \textbf{TN} & \textbf{FN} \\
\midrule
Qwen3-Omni-30B-A3B \cite{qwen3Omni} & 18.2 & \textbf{8.1}  & \underline{8.8} & 64.9 \\
R1AQA \cite{R1AQA}             & \textbf{24.3} & 54.7 & 2.7  & \textbf{18.3} \\
Audio Flamingo-3 \cite{flamingo3}   & 10.1 & \underline{16.2} & \textbf{16.9} & 56.8 \\
TimeAudio \cite{wang2025timeaudio}+Qwen3-8B \cite{qwen3} & \underline{20.2} & 41.9 & 6.8  & \underline{31.1} \\
AUDSEMTHINKER-QA GRPO \cite{audsemthinker}     & 10.1 & 29.8 & \textbf{16.9} & 43.2 \\
\bottomrule
\end{tabular}%
}
\vspace{-10pt}
\end{table}

\subsection{Bottleneck in Polyphonic Scenarios}
As shown in Table \ref{tab:polybench_results}, the five tasks exhibit a clear difficulty stratification: Duration, Concurrency, and Classification are relatively stable overall, whereas Counting and Detection are notably more challenging and suffer more pronounced performance drops. In particular, accuracies are generally low on Counting and Detection, two tasks that heavily rely on polyphonic parsing and temporal-structure judgments, indicating that unstable perceptual evidence under polyphony is further amplified in structured decisions such as cardinality estimation and interval-level localization.
Concretely, even the best-performing model, Qwen3-Omni-30B-A3B, achieves only 57.5\% accuracy on Counting and 63.4\% on Detection. For most other models, performance on these two tasks declines further: For example, Audio Flamingo 3 attains only 37.2\% on Detection, AUDSEMTHINKER-QA GRPO also reaches 37.2\% on Detection, and R1AQA obtains 30.1\% on Counting. 

Results on Concurrency and Classification are more competitive: Qwen3-Omni achieves 83.1\% on Concurrency and 77.9\% on Classification, while AUDSEMTHINKER-QA GRPO also attains a relatively high 70.3\% accuracy on Classification. This contrast suggests that when the task primarily requires deciding whether overlap exists or identifying the concurrent category under an anchor-event constraint, some models can still maintain relatively stable decisions. However, when the task further demands the deduplicated counting of events in a polyphonic scene or interval-level identification of the first overlap, errors increase substantially. In particular, on the Detection subset, TimeAudio+Qwen3-8B achieves the second-best accuracy (51.7\%), suggesting that a cascade paradigm may be advantageous for interval-level localization: the first-stage temporal localization or event perception provides more structured evidence for the second-stage pure text-based decision.

These observations are consistent with prior comprehensive evaluations: in the MMAU-Pro skill profile, Quantitative Reasoning and Temporal Event Reasoning are often among the weaker capability dimensions; similarly, in fine-grained temporal reasoning benchmarks such as TREA, Counting and Duration estimation frequently fail to reach stable performance. PolyBench further indicates that under polyphonic concurrency, these weaknesses are more readily triggered and amplified.


\subsection{Robustness Divergence in Compositional Reasoning}

Despite the substantial perceptual interference introduced by polyphonic overlap, models with different architectures exhibit clear performance stratification on compositional reasoning tasks such as Concurrency and Classification. This stratification reflects their differing abilities to maintain a stable decision process that links acoustic event evidence, relational constraints, and multiple-choice selection under polyphonic conditions. As shown in Table 1, on the Concurrency task, Qwen3-Omni-30B-A3B achieves 83.1\% accuracy, substantially outperforms Audio Flamingo 3 (66.9\%), AUDSEMTHINKER-QA GRPO (53.4\%), TimeAudio+Qwen3-8B (51.4\%), and R1AQA (42.6\%). 
This result indicates that, for polyphonic discrimination centered on the presence of overlap, Qwen3-Omni-30B-A3B can more reliably detect concurrent-event cues and therefore exhibit stronger robustness overall. 

A similar pattern emerges for Classification, which additionally requires identifying the category of the concurrent event. Qwen3-Omni-30B-A3B reaches 77.9\% accuracy, while AUDSEMTHINKER-QA GRPO attains a comparably high 70.3\%, clearly exceeding Audio Flamingo 3 (50.3\%) and TimeAudio+Qwen3-8B (50.3\%); R1AQA drops to 33.1\%. These results suggest that AUDSEMTHINKER-QA GRPO, which is explicitly fine-tuned for QA and further strengthened via GRPO-based reinforcement learning, exhibits improved robustness in audio-grounded logical judgment and option prediction. More broadly, when the task progresses from overlap existence detection to concurrent-category identification under an anchor-event constraint (including the suppression of distractors), the reliability of audio–text joint representations and the stability of option alignment become decisive factors that directly shape the attainable performance ceiling.

From a mechanistic perspective, the observed discrepancies are plausibly driven by two key capabilities under polyphony. The first is cross-modal alignment and representation stability, namely maintaining consistent category evidence despite acoustic masking. The second is instruction-following and discrete decision stability, namely consistently mapping the inferred concurrency relation onto the candidate options under MCQA constraints while resisting spurious distractors. Using Classification as an example, the model must not only perceive the presence of concurrent events but also perform category discrimination and distractor elimination within the anchor-event window. This places stronger demands on end-to-end stability spanning polyphonic cue extraction, semantic alignment, and relationally constrained selection. In contrast, if the underlying event evidence becomes unstable under polyphony (e.g., the concurrent event is missed or confused), even models with non-trivial language-side inference may fail to maintain consistent option selection, resulting in a decrease in accuracy.

Although frontier models demonstrate stronger robustness on Concurrency and Classification, they do not reach comparable levels on Counting and Detection, which rely more heavily on counting unique event categories and interval-level temporal localization. This further indicates that compositional reasoning in polyphonic audio, as targeted by PolyBench, remains constrained by perceptual reliability and the stability of temporal-structure parsing. Consequently, improving compositional reasoning under polyphonic conditions requires, first, strengthening low-level event perception and temporal-structure modeling to mitigate masking-induced evidence instability, and second, enhancing cross-modal alignment and task-constrained learning so that higher-level decisions are more reliably grounded in robust acoustic evidence and can generalize to more demanding compositional tasks.

\subsection{Performance Illustration and Shortcut Learning in Concurrency}

The results in Tables \ref{tab:polybench_results} and \ref{tab:concurrency_mixed_mono_confusion} indicate that current LALMs exhibit a pronounced performance illusion and shortcut learning on the Concurrency task. For example, R1AQA achieves 93.7\% F1 and 90.4\% ACC under the pure-polyphonic setting (Table \ref{tab:polybench_results}), appearing nearly perfect. However, under the equal-duration mixed monophonic and polyphonic evaluation constructed by jointly incorporating the AudioTime dataset \cite{AudioTime}
 (Table \ref{tab:concurrency_mixed_mono_confusion}), the overall correctness of all models drops substantially, suggesting that these LALMs have not yet developed a generalizable auditory discrimination capability for concurrency. Two systematic biases emerge: (i) a strong ``Yes" bias (e.g., R1AQA; FP = 54.7\%, TN = 2.7\%), indicating heavy reliance on task or prompt priors rather than audio evidence; and (ii) an overly conservative ``No" bias (e.g., Qwen3-Omni-30B-A3B; FP = 8.1\%, FN = 64.9\%), with ``Yes" predicted only when overlap evidence is highly salient.

In summary, the high scores in Table \ref{tab:polybench_results} likely reflect label-distribution bias rather than true overlap understanding, whereas the equal-duration mixed evaluation in Table \ref{tab:concurrency_mixed_mono_confusion} separates prior-driven answers from audio-evidence reasoning and reveals deficiencies in overlap perception.

\section{Conclusion}

We present PolyBench, a benchmark specifically designed to evaluate compositional reasoning in polyphonic audio with LALMs. PolyBench is composed of five structured MCQA tasks built on diverse real-world audio recordings. Extensive experiments reveal that current LALMs exhibit clear performance degradation under polyphonic conditions, particularly in tasks requiring temporal overlap reasoning, highlighting a fundamental gap in concurrent event perception. 

\section{Generative AI Use Disclosure}
We use generative AI tools to polish the manuscript, e.g., correcting the grammar.

\bibliographystyle{IEEEtran}
\bibliography{mybib}

\end{document}